# ChurnBench: A Drift-Aware Benchmark Demonstrating That Refresh Scheduling, Not Cache Age, Governs Staleness in Agentic AI

Vivek Kumar Singh
*Independent Researcher*
McKinney, TX, USA
vivekksingh.nov12@gmail.com
ORCID 0009-0002-9350-3207

Preeti Priyam
*Independent Researcher*
McKinney, TX, USA
preetipriyam12@gmail.com
ORCID 0009-0000-9423-741X

Gautam Bhowmick
*Independent Researcher*
Chicago, IL, USA
bhowmick.gautam@gmail.com
ORCID 0009-0001-9424-7826

***Abstract*—In production, agentic systems answer questions over data that lives in several places and keeps changing: licenses are reassigned, users offboarded, prices changed, contracts renewed. Existing retrieval benchmarks freeze the data, so they cannot ask whether an agent's answer is still true, only whether it found the right passage. We present ChurnBench, an open-source benchmark that generates a four-source enterprise data fabric as a timeline rather than a snapshot. Every change is written to an append-only ground-truth ledger, and gold answers are computed from that ledger, never from the live stores. An answer that was correct when its data was retrieved but wrong when evaluated is therefore detected and labeled a freshness error, distinct from a reasoning error; we validate this by resolving ground truth at both timestamps for every case reported. Using the instrument, we find that when a system refreshes on a schedule, cache age does not predict staleness. Across cache ages of 1, 14, and 28 days, freshness errors were 7, 4, and 4, because scheduled refresh bounds staleness by time-to-live, and no TTL lapse was observed in any window. A controlled ablation confirms the mechanism: disabling tiered refresh raises freshness errors from 4 to 45 at 28 days and leaves them identical at one day. The variable a drift benchmark should sweep is therefore TTL configuration against each entity's rate of change, not drift-window length. ChurnBench, the evaluation harness, and all per-error data are released open source at https://github.com/vsingh45/churnbench.**

## I. Introduction

An agent that answers enterprise questions has two jobs, and the literature has focused on one of them. The first is reasoning: break down the question, call the right tools, and compose an answer. The second is quieter, and in production it is harder: make sure the data underneath that reasoning comes from the right source, is joined correctly across systems that share neither keys nor vocabularies, and is still true. In our experience building software asset management platforms over warehouse, document-store, software-as-a-service (SaaS), and contract-document sources, the second job is where most failures come from. The agent is rarely wrong because the model could not reason. It is wrong because a cache served last month's license assignments, or an index built in January answered a question about March.

The problem is hard to study because staleness is invisible in existing evaluations. Benchmarks in the retrieval-augmented generation (RAG) and agent literature freeze a corpus, generate questions against it, and grade retrieval or final answers [1], [2], [3]. A frozen corpus cannot produce a stale answer at all. So the failure mode is not just unmeasured; it cannot be measured under the usual evaluation design. Recent work has started to address temporal drift for document corpora and for agent memory [4], [5], but the enterprise version of the problem is architectural, not textual. Here the staleness lives in caches, materialized aggregates, and vector indexes sitting over databases and rate-limited interfaces, not in a corpus of prose. A rule that replaces an outdated fact in a memory store does nothing for a materialized aggregate that was computed last Tuesday.

We built ChurnBench to close that gap. The core design choice is that the world is generated as a timeline, not a snapshot. A deterministic simulator plays out enterprise activity day by day and writes every change to an append-only ledger, and the live stores an agent queries are just folds of that ledger at a chosen timestamp. Because ground truth comes from the ledger and never from the stores, an answer that was correct when its facts were retrieved but wrong when it was evaluated can be detected automatically and labeled a freshness error, separate from an answer that was never correct at all.

With the instrument in hand, we used it to test the idea that motivated it: that staleness grows with the age of the cache an agent reads from. For the systems most likely to be deployed, that idea is wrong, and the reason is worth understanding. Systems that refresh on a schedule bound their staleness by time-to-live (TTL), not by how long ago the cache was built. An entity with a one-day TTL is refreshed again and again during any window longer than a day, so it arrives at evaluation about one day old whether the cache was built one day or 28 days earlier. Cache age, the natural variable to sweep in a drift benchmark, never actually reaches the retrieval layer.

We report this negative result, the mechanism behind it, and a controlled ablation that isolates the mechanism by turning scheduled refresh off while holding everything else fixed. We report it because a benchmark whose main variable cannot move its outcome is a design error, and one that others building drift benchmarks are likely to repeat. The fix is to sweep the TTL configuration against each entity's rate of change, and to record per-entity refresh timestamps rather than counts of expiry events.

This paper makes three contributions. *First*, ChurnBench [14]: an open-source benchmark that generates a four-source enterprise data fabric as a timeline with an append-only ground-truth ledger, supporting frozen-at-T evaluation of agentic systems over drifting data, with programmatically resolved gold answers and no model in the grading path. *Second*, a verifiable freshness-error protocol that separates staleness from reasoning failure by re-resolving ground truth at the answer's effective retrieval time, applied and confirmed for every case reported here. *Third*, the negative empirical finding described above, together with a fully matched controlled ablation establishing that tiered refresh scheduling, not cache age, governs observed staleness, and an attribution analysis showing that staleness exposure is governed by the product of tier width and entity mutation rate.

We are clear about scope. This paper does not claim that any architecture beats another. We describe one grounding architecture in Section III because it is the system we measure, but no baseline comparison was run under the final design, and no performance claim should be read into the accuracy numbers in Section VI. What we claim is an instrument, its validation, and what it revealed when pointed at a system that behaves the way production systems do.

## II. Related Work

### A. Retrieval Benchmarks and Their Corpora

Retrieval-augmented generation has moved from single-shot retrieve-then-generate toward modular and agentic variants that break questions down, issue sub-queries, and iterate [6]. The benchmarks underneath this progress have stayed static and single-source. BEIR [1] gathers many retrieval tasks, but each is a fixed snapshot; MS MARCO [7] and the multi-hop question sets [8], [9] come from web and encyclopedia crawls captured at one moment. Practitioner reports note that no public benchmark reflects enterprise deployments, since most use web data alone. WixQA [10] moves to an enterprise help-center corpus with a pinned knowledge-base snapshot, which sharpens grading but still keeps the world frozen. ChurnBench differs on both counts: several heterogeneous sources across three modalities, and a world that changes between when the index is built and when the agent is evaluated.

### B. Enterprise Agent Benchmarks

Enterprise agent benchmarks have started to appear. AgentArch [2] evaluates eighteen agent configurations across orchestration, prompting, and memory on enterprise tasks, and finds that even strong models reach only modest success on the harder tier. A supervisor-worker design over separate relational and document stores has been offered as an open-source enterprise pattern [11], though without a published evaluation. Execution-grounded text-to-SQL benchmarks [3] led us to grade by execution rather than by rubric, though they are single-step and have no staging layer. All of these evaluate the agent while treating the data underneath as a fixed snapshot. None injects drift, and none measures freshness. We do the complementary experiment: hold the agent fixed and vary the data layer beneath it.

### C. Temporal Drift and Staleness

A recent line of work treats knowledge drift as a first-class concern. Liu et al. [4] build a benchmark of real-world events from time-stamped evidence and show that both RAG and learning-based adaptation struggle to stay temporally consistent as facts change. MemStrata [5] defines a stale-fact-error rate over changing text facts, reports that retrieval systems serve outdated values fairly often, and removes the error class with a deterministic supersession rule over a bi-temporal ledger.

We share the bi-temporal-ledger idea with MemStrata but move the problem. Their staleness lives in an agent's memory over prose facts in one corpus; ours lives in the grounding layer, spread across caches, materialized aggregates, and vector indexes over databases, interfaces, and documents. The two approaches complement rather than compete: a supersession rule keeps remembered facts current, but it cannot refresh a materialized aggregate or expire a cache tier, because those have no fact-level identity to supersede. To our knowledge, no prior benchmark sits at the intersection we target: multi-source structured and unstructured enterprise data, controlled drift, execution-based grading, and per-task cost. Each of these exists on its own in prior work; what is new here is the combination, and the instrument that measures within it.

## III. The System Under Measurement

ChurnBench measures agentic systems that answer questions over a set of heterogeneous sources. To exercise the benchmark, we built one such system: a grounding layer placed between the sources and the agent runtime. We describe it because it is what we measure, not because we claim it is better than the alternatives. Fig. 1 shows the full pipeline, from timeline generation through projection, retrieval, and scoring.

### A. Semantic Model

A declarative registry records, for each entity class, its source, its location in the staged store, a freshness tier, a TTL, the measures it can answer, and the time of its last refresh. In the software asset management setup, license assignments and user status are hot entities with a one-day TTL; prices and cost-center membership are warm at seven days; contract terms and vendor dimensions are cold at thirty. Consumption facts are marked live and are never staged.

The registry is data, not prompt text, which matters for two reasons. It makes routing decisions inspectable and deterministic, and it makes coverage explicit and auditable: a measure is either in the registry or it is not. Section VI-D reports what happens to accuracy when it is not.

### B. Source-Aware Routing

Given a retrieval need, a rule-based router picks among five paths without calling a model: a templated query against the staged store; a federated template that runs one staged query and one live warehouse query and joins the results in the harness; a live warehouse query for entities marked live; a call to the origin interface for needs that must be live; and a vector lookup over the document index. Because routing is deterministic, the path taken for any question is reproducible and recorded in the trace, along with the last-refresh time of every entity read. That record is what makes the freshness-error protocol of Section IV-E possible.

### C. Tiered Refresh

Dedicated pipelines refresh each entity class on a schedule set by its tier. The harness moves day by day from cache construction to evaluation and, each day, refreshes any entity whose TTL has lapsed, stamping a new last-refresh time. The idea is standard: fast-changing entities refresh often, stable ones refresh rarely, keeping staleness low where it matters without paying for needless refreshes. This is exactly the mechanism the experiments in Section VI test, and turning it off is the single variable we change in the controlled comparison.

How ChurnBench tells a stale answer from a wrong one
DATA PATH
what the agent actually sees when it answers
Timeline Simulator
plays out the world
Ground-Truth Ledger
every change, in order
56,370 events
fold
at T
Four Sources
• SQL warehouse
• document store
• rate-limited API
• contract corpus
Grounding Layer
decides where each answer comes from
semantic model — tier · TTL · origin
tiered refresh — 1 d / 7 d / 30 d
router — 6 possible routes, rule-based
Agent Runtime
LangGraph
Answer
+ trace
$T_{eff}$
when its data was last true
TRUTH PATH
what the answer is checked against
Gold Resolver
reads the ledger — no model
asked twice, for two moments in time
gold at T — is it true now?
gold at $T_{eff}$ — was it true when read?
Verdict
correct — matches gold now
freshness error — true then, not now
reasoning error — never true
A wrong answer counts as a freshness error only if it was true at the moment its data was retrieved.
Everything above is what the agent can see; everything below decides whether it was right.

Fig. 1. ChurnBench end to end. A seeded simulator writes every mutation to an append-only ledger; the four live sources are folds of that ledger at the evaluation timestamp, and gold answers are resolved from the ledger rather than from the stores. The grounding layer registers each entity class with a freshness tier and refreshes it on schedule, while the router records, for every retrieved fact, which source served it and when that entity was last refreshed. Scoring resolves gold twice, at the evaluation time and at the effective retrieval time, which is what separates a stale answer from a wrong one.

## IV. ChurnBench

### A. The Simulated Fabric

The fabric has four sources chosen to match the modality mix of a real enterprise stack. A PostgreSQL warehouse holds historical facts in a star-adjacent schema: license purchases, consumption events, and vendor, product, and cost-center dimensions. A MongoDB store holds current operational state: users, active licenses, assignments, and entitlements. A mock REST interface stands in for a SaaS system, with 350 ms of added latency and a limit of sixty requests per minute, so that designs relying on live per-entity calls pay a realistic price. A set of synthetic contract documents is rendered from the ledger, so renewal terms and cancellation windows exist only in prose and can be reached only through the document index.

The domain is software asset management. We chose it because drift is built into it: reassignment, offboarding, renewal, and repricing are the domain's day-to-day activity, and the same estate naturally spans structured usage data and unstructured contract terms.

### B. Timeline Generation and the Ledger

A deterministic simulator plays out a span of days. Each day it draws event counts from independent Poisson streams for hires, offboardings, license reassignments, price changes, contract renewals, and cost-center moves, plus a dense consumption stream, and writes every event to the ledger with a monotonic sequence number. Identical seeds produce identical ledgers, which we verify by test. A second parameter concentrates a configurable share of changes on the top ten percent of entities, reflecting the fact that a few products and users change far more than the rest.

The stream rates are calibrated against published statistics rather than picked freely. National labor-turnover data [13] sets the offboarding rate for an estate of this size, and industry software-estate reports set how common idle licenses are and how often contracts renew. Reassignment and repricing rates have no public source and are set by practitioner judgment; the experiments below span a wide range around the defaults, so no conclusion rests on any one calibration choice.

The live stores are projections. Folding the ledger up to a timestamp rebuilds the warehouse, the operational store, the interface state, and the document corpus exactly as they stood at that moment. Projection is idempotent, and we test it. Gold answers are computed from the ledger by a non-model resolver, never from the live stores, so no grader model's errors can leak into the measurement and no bug in the projection layer can silently change what counts as correct.

### *C. Frozen-at-T Evaluation*

A run fixes two timestamps. The cache and indexes are built at $T'$; the questions are asked, and the world stands, at $T$. The gap between them is the cache age. Freezing the world during a run gives up one bit of realism, drift continuing mid-run, in exchange for exact reproducibility, which we think is the right trade for a first instrument.

### *D. Tasks*

Tasks are generated from parameterized templates over four executive intents: spend visibility, savings opportunity, criticality, and utilization. Three difficulty tiers cover single-source questions, cross-source questions that reconcile the relational and operational stores, and cross-modality questions that combine structured facts with contract prose. The run reported here uses 180 tasks, split 81, 63, and 36 across the tiers and 50, 38, 48, and 44 across the intents.

We guard against degenerate answers explicitly. A trivial responder that always answers zero or none scores 1.7 percent on the generated set, enforced by a committed test, so accuracy cannot be gamed by refusing to answer. Templates whose gold answer is always zero were replaced during development for the same reason.

Two kinds of question are outside the reach of a current-state grounding architecture, and we keep them rather than remove them. Event-history questions, such as how many licenses were reclaimed or how many reassignments happened in a window, need an audit trail the standard fabric does not keep; current-state stores record what is, not what changed. Per-license cost totals need a per-license price table the warehouse does not have, since prices attach to purchase batches rather than individual seats. A benchmark with only answerable questions would overstate what any system can do, so these stay in and their failures are reported.

### *E. Freshness Error*

Four measures are recorded per task: accuracy against ledger-derived gold; cost in United States dollars with exact token attribution; end-to-end latency; and freshness error. A task is a *freshness error* if and only if the answer is wrong against gold at $T$ and correct against gold at $T_{eff}$, the effective retrieval time of the facts used. $T_{eff}$ is derived from the trace rather than assumed: live routes contribute $T$, and staged routes contribute the last-refresh timestamp of the entity actually read, with the minimum taken when several entities contribute. An answer that matches gold at neither timestamp was never true of any world and is a *reasoning error*, scoring against accuracy alone.

This definition is the paper's measurement contribution, and it can be checked rather than merely asserted. For every freshness error below, we resolved the ledger at both timestamps and confirmed both conditions hold. The check also catches a case that would otherwise be invisible: a wrong answer that happens to match some earlier world state satisfies the first condition but not the second, and is correctly excluded.

## V. Experimental Setup

All runs in Table I use one model, nvidia/nemotron-3-ultra-550b-a55b, served through NVIDIA NIM at temperature zero with the model's reasoning mode disabled, and one agent framework, LangGraph. The three-window sweep in Table III was run earlier with reasoning mode enabled; we report it separately for that reason, and note in Section VI-B that the freshness-error count at the 28-day window is the same under both settings. The fabric runs under Docker with PostgreSQL 16 and MongoDB 7; the document index uses a local sentence-transformer model, so embedding cost is zero though tokens are still counted. The implementation is about 7,500 lines of Python with 316 tests. A 180-task run costs about 1.50 US dollars in inference at list prices with reasoning on, and about 0.50 dollars with it off, averaging 0.008 and 0.003 dollars per task.

The ledger for these runs holds 56,370 events, most of them 54,000 consumption events, plus 717 hires, 605 license reassignments, 237 offboardings, 200 purchases, 200 assignments, 170 unassignments, 112 cost-center moves, 77 price changes, 27 contract renewals, and 25 contract signings. That contract renewals are rare while consumption events are dense matters for the attribution analysis in Section VI-C.

Four runs hit interface instability that made some tasks fail after exhausting retries, including the ablation run that produces the paper's largest effect. In each case the affected tasks were re-run on their own under the same configuration with a hardened retry policy, the other results were kept byte-for-byte and checked against a pre-merge backup, and the change was recorded in the result file. Where an original run predated a configuration change, the retry was run from the original commit in a separate worktree, so all tasks in a result set share one configuration. Every cell of Table I is free of unresolved failures. Section VII reports what the retries did and did not change.

No baseline comparison was run under the final design. Naive tool-calling, single-index retrieval, and hierarchical decomposition arms are in the released code and were used during development, but they were not evaluated on the 180-task set under the configuration reported here. The accuracy numbers in Section VI describe one implementation under two refresh settings; they are not evidence that any architecture beats another, and we draw no such conclusion.

Every number in this paper can be regenerated from the public repository [14]. The ledger for the reported runs is committed, along with the generation command and seed that produce it; identical seeds yield identical ledgers. Each result file records the commit hash, model string, T′, T, seed, and task-set hash it was produced under, and every retried task carries a provenance record naming the retry configuration. The supplementary data include, for each freshness error, the effective retrieval time and the gold values at both timestamps, so the definitional check of Section IV-E can be re-run from committed files alone without re-executing any model call. Per-statistic supplementary files back each table individually, including per-entity cache ages, the sweep summary, task-set composition, repository statistics, and a provenance row for every run; a reviewer guide maps each number in this paper to the file and command that reproduces it. An extended architecture description is maintained alongside the code in ARCHITECTURE.md [14].

## VI. Results

### *A. Tiered Refresh Governs Staleness; Cache Age Does Not*

The primary result is an interaction. We ran the system in two settings, tiered refresh on and tiered refresh off so the cache is built once and never refreshed, at two cache ages, holding the model, task set, seed, fabric, prompts, reasoning mode, and everything else fixed. All four cells ran from behaviorally identical code with reasoning mode off and no unresolved failures, so the only things that change are the refresh setting and the cache age. Table I reports freshness errors in each cell.

TABLE I. FRESHNESS ERRORS BY REFRESH CONFIGURATION AND CACHE AGE (180 TASKS PER CELL)

| Refresh configuration | 1-day cache | 28-day cache |
|---|---|---|
| Tiered refresh enabled | 7 | 4 |
| Tiered refresh disabled | 7 | **45** |

All four cells share one configuration and differ only in the refresh setting. Every reported freshness error satisfies the definitional check of Section IV-E.

Disabling tiered refresh raises freshness errors elevenfold at a 28-day cache age and changes them not at all at one day: both configurations produce exactly seven, from the same seven tasks and the same entity classes. The asymmetry is the finding. Over a single day a frozen cache and a scheduled one are not merely similar but identical, because nothing has had time to drift and there is therefore nothing for refresh to prevent. Over 28 days the frozen cache accumulates staleness while the scheduled one does not accumulate it at all. An effect that is exactly zero at short windows and large at long ones is what a mechanism bounded by time-to-live predicts, and it is not what a mechanism driven by cache age would predict, since cache age differs by a factor of 28 across the two columns in both rows.

The mechanism is visible directly in the per-entity refresh timestamps in the traces, shown in Table II and Fig. 2. With tiering on, entities with a one-day TTL have a last-refresh time within one day of evaluation, whether the cache was built 1 or 28 days earlier. Entities with a seven-day TTL show ages of one to six days. With tiering off, every entity's last-refresh time equals the cache-build time, 29 days stale.

TABLE II. OBSERVED CACHE AGE PER ENTITY CLASS AT A 28-DAY CACHE AGE

| Entity class | TTL (d) | Tiered (d) | Untiered (d) |
|---|---|---|---|
| assignments | 1 | 1 | 29 |
| user status | 1 | 1 | 29 |
| prices | 7 | 5 | 29 |
| contract terms | 30 | 29 | 29 |

Contract terms are unchanged between configurations because a 30-day time-to-live never lapses within a 28-day window.

### *B. Why the Cache-Age Sweep Fails*

With tiered refresh on, the setting a production system would run, we swept cache age across 1, 14, and 28 days. Table III reports the result. The three windows are not configuration-matched: the 14-day run predates a model-configuration change and ran with the model's reasoning mode enabled, while the other two ran with it disabled. We report the sweep with that caveat rather than omit the middle window, because the freshness-error count at the 28-day window is 4 under either setting, so the flat trend does not depend on the configuration. Freshness errors are flat and non-monotonic, and accuracy varies by a few points with no trend. On its own, this is an uninformative null. Read with Table II, it is a mechanism.

TABLE III. CACHE-AGE SWEEP WITH TIERED REFRESH ENABLED

| Cache age | Freshness errors | Accuracy | TTL lapses |
|---|---|---|---|
| 1 day | 7 | 67.2% | 0 |
| 14 days | 4 | 69.4% | 0 |
| 28 days | 4 | 65.0% | 0 |

The three windows are not configuration-matched: the 14-day run predates a model-configuration change and executed with the model's reasoning mode enabled, while the 1-day and 28-day runs executed with it disabled. The freshness-error count at 28 days is 4 under either setting, so the flat trend does not depend on the choice. No time-to-live lapse was observed at query time in any window: scheduled refresh renewed every entity before its tier expired.

No TTL lapse was recorded at query time in any window. The scheduler moves forward from cache build to evaluation and refreshes any entity whose tier has expired, so an entity with a one-day tier refreshes over and over during the walk and always arrives within one day of evaluation. Cache age, the variable the sweep was meant to change, never reaches the retrieval layer for any entity whose TTL is shorter than the window. Only contract terms, on a thirty-day tier, keep the original build time, and only because thirty is longer than the widest window we tested. Fig. 2 makes this explicit.

This is the design error we want to document. A drift benchmark that sweeps the gap between index build and evaluation quietly assumes the index is never maintained. For any system that refreshes on a schedule, that assumption is false and the sweep measures nothing. A wider window does not fix it; it only moves the point at which the longest tier also stops mattering. The fix is to change the swept variable to the TTL configuration itself, holding cache age fixed.

**(a) TIERED REFRESH ENABLED — cache age is a bounded sawtooth**

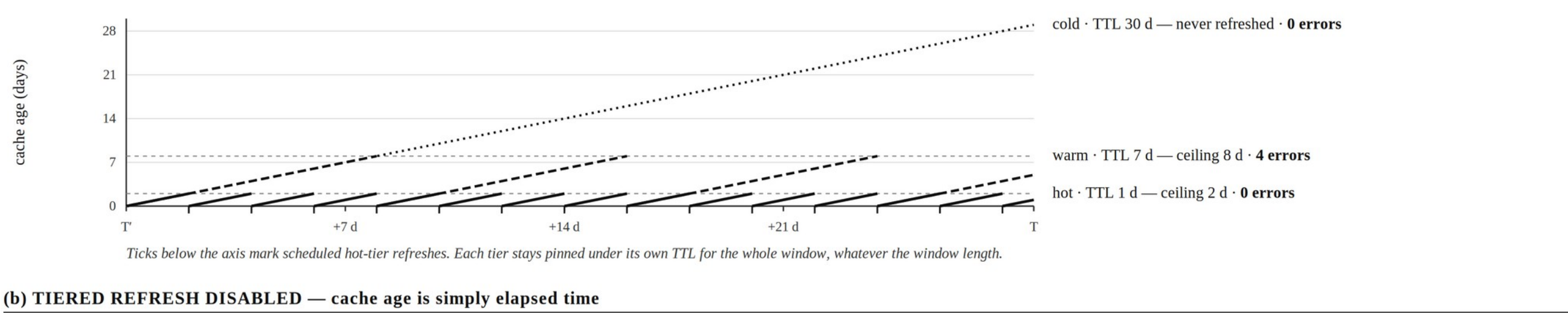

**(b) TIERED REFRESH DISABLED — cache age is simply elapsed time**

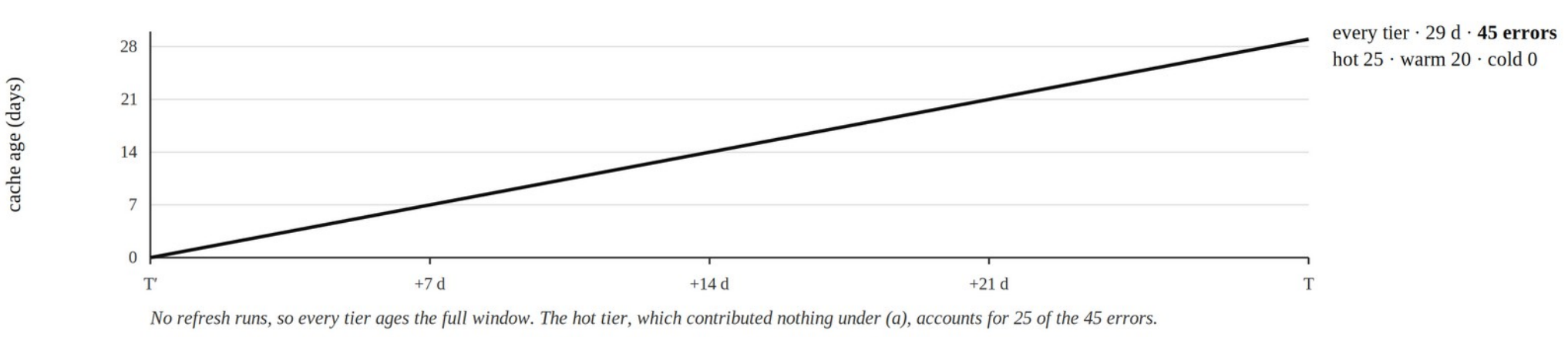

*Same 180 tasks, same model, same window; the only difference is whether the daily refresh walk executes.*

Fig. 2. Why cache age cannot move the dependent variable. Each curve is one entity class's cache age across the walk from T′ to T. With tiered refresh enabled (a), age resets on every scheduled refresh and stays bounded by the tier's own TTL for the whole window, so the window length is irrelevant; only the 30-day tier, whose TTL never lapses inside a 28-day window, climbs with elapsed time, and it produces no errors because contract terms rarely change. With refresh disabled (b), every tier ages linearly and the hot tier, which contributed nothing under (a), accounts for 25 of the 45 errors.

### *C. Where Staleness Actually Lands*

Attribution sharpens the picture. With tiering on, fourteen of the fifteen freshness errors across the three windows came from one entity class, prices, on the seven-day tier, reached through templated queries against the staged store. Two measures account for those fourteen: nine from cost-center spend totals and five from finding the highest-spending cost center. The fifteenth, at the one-day window, came from contract terms and is discussed below. One-day-tier entities produced none, not because they are immune but because a one-day staleness window rarely spans a change. Thirty-day-tier entities also produced none, despite being genuinely stale, because contract terms change rarely: twenty-seven renewals across 56,370 events.

TABLE IV. FRESHNESS ERRORS BY ENTITY CLASS AT A 28-DAY CACHE AGE, MATCHED PAIR

| Entity class | Tier | TTL (d) | Tiered refresh | Untiered |
|---|---|---|---|---|
| user status | hot | 1 | 0 | 15 |
| assignments | hot | 1 | 0 | 10 |
| prices | warm | 7 | 4 | 20 |
| contract terms | cold | 30 | 0 | 0 |
| **Total** | | | **4** | **45** |

Under tiered refresh, only the warm tier produces staleness. Removing refresh exposes the hot tier, which changes fastest, while the cold tier stays at zero because its entities rarely change.

With tiering disabled at the 28-day window the distribution changes qualitatively. Of the forty-five freshness errors, fifteen come from user status and ten from assignments, the one-day-tier entities that produced zero errors under tiering, with twenty from prices. The one-day window behaves differently again: both configurations produce the same seven errors, six from prices and one from contract terms. That single cold-tier error is the only one anywhere in our data, and it arises because a contract mutation happened to fall inside the window; a thirty-day tier is stale by construction, but contract terms mutate so rarely that staleness almost never coincides with a query.

The generalization is that staleness exposure is governed by the product of tier width and entity mutation rate, and that scheduled refresh is what keeps that product small. Neither factor alone predicts where errors appear: the widest tier produced none because its entities barely change, and the fastest-changing entities produced none until their tier was removed. The design implication for practitioners is that tiers should be assigned from measured mutation rates rather than from intuitions about which data feels important.

One methodological artifact deserves note, because it would mislead anyone instrumenting a similar system. The count of time-to-live lapse events reads zero in both configurations, for opposite reasons. Under tiering, no tier expires before refresh renews it. Without tiering, the staleness check is skipped entirely and never evaluates expiry at all. A metric that reports identical values for opposite behaviors is a measurement trap. Per-entity refresh timestamps, not lapse counts, are the reliable instrument, and we recommend that any system in this class record them.

### *D. Accuracy by Task Tier, and the Effect of Coverage*

Table V breaks accuracy down by task tier for the matched pair. The tiers discriminate as intended: single-source questions are answered far more often than cross-source ones, and disabling refresh costs the most on the tiers that touch stageable entities. Tier 3, which reaches contract prose through the document index, is nearly unchanged between settings, because the document index is rebuilt at T′ in both.

TABLE V. ACCURACY BY TASK TIER AT A 28-DAY CACHE AGE, MATCHED PAIR (N PER TIER: 81 / 63 / 36)

| Task tier | Tiered refresh | Untiered |
|---|---|---|
| 1 — single-source | 80.2% | 42.0% |
| 2 — cross-source | 47.6% | 17.5% |
| 3 — cross-modality | 55.6% | 55.6% |
| **All (n = 180)** | **65.0%** | **36.7%** |

Tier-3 accuracy is identical across settings because those tasks are served from the document index, which both configurations build at the same time.

Accuracy in Tables I and III is dominated by a property of the reference implementation rather than of the architecture, and we report it with that caveat attached. The 180-task set exercises 22 distinct measures, of which 9 are registered in the semantic model and 13 are not. One hundred and nine of the 180 tasks target unregistered measures and fall through to model-generated SQL against the staged schema. Between 49 and 53 of the 51 to 69 reasoning errors per run come from exactly those tasks.

Reported accuracy therefore largely measures registry coverage. This is a limitation of the implementation, not a finding about the architecture, and it should not be read as one. It is also, we note, the failure mode the semantic model makes visible: coverage is an auditable property of a registry, so a gap is diagnosable by inspection rather than only by degraded answers. Expanding coverage is engineering rather than research, and we leave it to future work.

## VII. Threats to Validity

*Construct validity.* A skeptic might say a freshness error is just an accuracy error with a timestamp. The point is the attribution. An answer that was true of an earlier world points at the grounding layer and is fixed by a refresh policy; an answer that was never true points at the model and is not. Blurring the two is how the problem stayed invisible. The definitional check makes the distinction concrete rather than rhetorical, and it is falsifiable: any claimed freshness error can be rejected by resolving the ledger again.

*Statistical power.* Freshness-error counts under tiering are small, between four and seven per 180-task run, and the flatness across the sweep rests on differences of at most three. So we do not rest the negative finding on the sweep alone. The mechanism evidence in Table II and the zero lapse counts are direct measurements, not subject to the same small-sample worry, and the ablation gap of 4 versus 45 is too large for sampling noise to explain.

*Instrument robustness under transport failure.* Four runs experienced interface instability causing between 4 and 24 tasks each to fail after exhausting retries, including the ablation run that produces the paper's largest effect, where four such failures were detected only on a later audit of every committed result file. All four were re-run in isolation under identical configuration, with the surviving results preserved byte-identically and verified against a pre-merge backup. In all four cases the freshness-error count was unchanged, while correct answers rose and parse failures fell to zero. The replication is the point: transport failures land in the reasoning and parse-failure categories rather than suppressing freshness errors, which follows from the definitional check requiring a substantive answer that matches an earlier world. A stub cannot match any world. We nonetheless recommend auditing for this class of failure by trace signature rather than by output string, since the latter missed a case here.

*Configuration comparability.* Earlier one-day and 28-day runs exist in the released repository that predate a model-configuration change. We retain them rather than delete them, and they are the source of the three-window sweep of Table III, which we report in its own configuration for that reason. The four cells of Table I were executed from behaviourally identical code under one configuration and differ in exactly one flag. Where a claim could depend on the configuration choice we report the value under both, as with the 28-day freshness count.

*Scope.* A single model, a single agent framework, a single domain, and synthetic data. Synthetic generation is what makes controlled drift and exact ground truth possible at all, but prevalence claims do not transfer to real estates, and the calibration of two mutation streams rests on practitioner judgment rather than published statistics. No baseline comparison was run, as stated in Section V, and no cross-provider replication was attempted.

## VIII. Discussion and Future Work

For practitioners, the knob to tune is the tier assignment, and the right basis for it is the measured rate of change. An entity that changes daily but sits on a weekly tier is where staleness will show up; an entity that changes once a year can sit on a long tier with no risk. Our data show exactly this: the only entity class that produced staleness under tiering was the one whose tier width and rate of change overlapped, and the widest tier was harmless because its entities barely changed. A team building a grounding layer can compute this from a change stream before picking a TTL.

For benchmark designers the lesson is sharper, and it is why we report the null result rather than shelve it. Drift-window length is the natural variable to sweep in a drift benchmark, and it is the wrong one for any system that maintains its indexes. We built such a sweep, ran it across three windows, and got a flat line; only the ablation recovered the signal. Future drift benchmarks in this space should sweep the TTL configuration against each entity's rate of change, should record per-entity refresh timestamps rather than counts of expiry events, and should check that the variable they sweep actually reaches the component under test before reading anything into a null.

The immediate next step is that sweep: hold cache age fixed, vary tier width across a grid, and measure the staleness-versus-cost trade directly, turning the guidance above from qualitative into quantitative. Three further directions follow. Letting drift continue during a run would remove the frozen-at-$T$ simplification. A second domain would test whether the tiering approach transfers or is specific to software asset management's pattern of change. And comparing grounding architectures on this instrument would answer the question this paper deliberately leaves open: whether staged, hierarchical, or live-only designs trade staleness against cost differently.

## IX. Conclusion

We built an instrument for measuring when an agentic system answers from data that has since changed, validated it by resolving ground truth at both the retrieval and evaluation timestamps for every reported case, and used it to test whether staleness grows with cache age. It does not. Under tiered refresh scheduling, staleness is bounded by TTL and is insensitive to how long ago the cache was built; a controlled ablation raises freshness errors from 4 to 45 by turning off the scheduler alone, and leaves them unchanged over a one-day window. Attribution shows the exposure is set by the product of tier width and entity rate of change. So the variable a drift benchmark should vary is the tier configuration, not the drift-window length. ChurnBench, the evaluation harness, and the complete per-error data are released open source [14] so that others can measure this rather than assume it.

## Acknowledgment

The authors used a generative AI system during manuscript preparation, limited to language editing and reviewed coding assistance. The method, benchmark design, experimental design, data, and analysis are the authors' own work.